\documentclass[]{acm_proc_article-sp}
\usepackage{url}

\usepackage{bigpage}
\usepackage{bcprules}
\usepackage{mathpartir}
\usepackage{listings}
\usepackage{mathtools}
\usepackage{amsfonts}
\usepackage{latexsym}
\usepackage{amssymb}
\usepackage{caption}

\newcommand{\ldb}{[\![}
\newcommand{\rdb}{]\!]}
\newcommand{\ldrb}{(\!(}
\newcommand{\rdrb}{)\!)}

\newcommand{\lpquote}{\ulcorner}
\newcommand{\rpquote}{\urcorner}

\newcommand{\id}[1]{\texttt{#1}}

\newcommand{\pzero}{\mathbin{0}}

\newcommand{\juxtap}{\mathbin{\id{|}}}
\newcommand{\concat}{\Rightarrow}

\newcommand{\scong}{\mathbin{\equiv}}
\newcommand{\nameeq}{\mathbin{\equiv_N}}
\newcommand{\alphaeq}{\mathbin{\equiv_{\alpha}}}
\newcommand{\names}[1]{\mathbin{\mathcal{N}(#1)}}
\newcommand{\freenames}[1]{\mathbin{\mathcal{FN}(#1)}}

\newcommand{\binpar}[2]{#1 \juxtap #2}
\newcommand{\outputp}[2]{#1 ! ( * #2 )}
\newcommand{\prefix}[3]{#1 ? ( #2 ) \concat #3}
\newcommand{\lift}[2]{#1 ! ( #2 )}
\newcommand{\quotep}[1]{@#1}
\newcommand{\dropn}[1]{*#1}

\newcommand{\xbangp}[2]{\int_{#2} #1}
\newcommand{\bangxp}[2]{\int^{#2} #1}

\newcommand{\substp}[2]{\id{\{} \quotep{#1} / \quotep{#2} \id{\}}}
\newcommand{\substn}[2]{\id{\{} #1 / #2 \id{\}}}

\newcommand{\psubstp}[2]{\widehat{\substp{#1}{#2}}}

\newcommand{\meaningof}[1]{\ldb #1 \rdb}
\newcommand{\pmeaningof}[1]{\ldb #1 \rdb}
\newcommand{\nmeaningof}[1]{\ldrb #1 \rdrb}

\newcommand{\Proc}{\mathbin{Proc}}
\newcommand{\QProc}{\quotep{\mathbin{Proc}}}

\newcommand{\defneqls}{\coloneqq}

\newcommand{\red}{\rightarrow}
\newcommand{\wred}{\Rightarrow}

\newcommand{\rel}[1]{\;{\mathcal #1}\;} 
\newcommand{\wbbisim}{\stackrel{\centerdot}{\approx}} 

\newcommand{\ptrue}{\mathbin{true}}

\newcommand{\pdropf}[1]{\rpquote #1 \lpquote}
\newcommand{\pquotep}[1]{\lpquote #1 \rpquote}
\newcommand{\plift}[2]{#1 ! ( #2 )}
\newcommand{\pprefix}[3]{\langle #1 ? #2 \rangle #3}
\newcommand{\pgfp}[2]{\textsf{rec} \; #1 \mathbin{.} #2}
\newcommand{\pquant}[3]{\forall #1 \mathbin{:} #2 \mathbin{.} #3}
\newcommand{\pquantuntyped}[2]{\forall #1 \mathbin{.} #2}
\newcommand{\riff}{\Leftrightarrow}

\newcommand{\PFormula}{\mathbin{PForm}}
\newcommand{\QFormula}{\mathbin{QForm}}
\newcommand{\PropVar}{\mathbin{\mathcal{V}}}

\newcommand{\bc}{\mathbin{\mathbf{::=}}}
\newcommand{\bm}{\mathbin{\mathbf\mid}}

\newlength{\ltext}
\newlength{\lmath}
\newlength{\cmath}
\newlength{\rmath}
\newlength{\rtext}

\newenvironment{grammar}{
  \[
  \begin{array}{l@{\quad}rcl@{\quad}l}
  \hspace{\ltext} & \hspace{\lmath} & \hspace{\cmath} & \hspace{\rmath} & \hspace{\rtext} \\
}{
  \end{array}\]
}

 \newtheorem{thm}{Theorem}[subsection]

 \newtheorem{defn}[thm]{Definition}

 \numberwithin{equation}{subsection}

\newcommand{\pic}{$\pi$-calculus}
\newcommand{\rhoc}{${\textsc{rho}}$-calculus}

\newcommand{\papertitle}{Policy as Types}

\begin{document}
\lstset{language=}

\setlength{\topmargin}{0in}
\setlength{\textheight}{8.5in}
\setlength{\parskip}{6pt}

\title{\papertitle}

\numberofauthors{3}
\author{
\alignauthor 
L.G. Meredith\\
  \affaddr{Biosimilarity, LLC}\\
  \email{\fontsize{8}{8}\selectfont lgreg.meredith@biosimilarity.com}\\
\alignauthor
Mike Stay\\
  \affaddr{Biosimilarity, LLC}\\
  \email{\fontsize{8}{8}\selectfont mike.stay@biosimilarity.com}\\
\alignauthor
Sophia Drossopoulou\\
  \affaddr{Imperial College, London}\\
  \email{\fontsize{8}{8}\selectfont s.drossopoulou@imperial.ac.uk}
}










\begin{abstract}
\normalsize{ 

  Drossopoulou and Noble \cite{Drossopoulou:2013:NCP:2489804.2489811}
  argue persuasively for the need for a means to express policy in
  object-capability-based systems. We investigate a practical means to
  realize their aim via the Curry-Howard isomorphism
  \cite{Abramsky:1992:PP:194588.194591}
  \cite{Krivine-TheCurryHowardCorre}. Specifically, we investigate
  representing policy as types in a behavioral type system for the
  \rhoc\; \cite{DBLP:conf/tgc/MeredithR05}, a reflective higher-order
  variant of the \pic\; \cite{milner91polyadicpi}.

}

\end{abstract}

\noindent
{\large \textbf{Submission to WPES}}\\
\rule{6.25in}{0.75pt}\\\\\\

\maketitle




\section{Introduction}

The object-capability (ocap) security model grew out of the realization
that good object-oriented programming practice leads to good security properties.
Separation of duties leads to separation of authority; information hiding
leads to integrity; message passing leads to authorization; 
and dependency injection to authority injection.  An
ocap-secure programming language enforces these patterns: the only way an
object can modify any state but its own is by sending messages on the object
references it possesses.  Authority can then be denied simply by not providing
the relevant object reference. Ocap languages do not provide ambient, undeniable 
authority such as the global variables in JavaScript,
static mutable fields in Java and C\#, or allowing access to arbitrary objects
through pointer arithmetic as in C and C++.

The electronic society is moving steadily towards the object capability model.
ECMAScript 5, the standard version of JavaScript adopted by all modern
browsers, introduced a new security API that enables the web browser to
be turned into an ocap-safe platform; Google's Caja project is an implementation
\cite{Caja:2013}.  Brendan Eich, CTO of Mozilla, wrote, ``In SpiderMonkey + Gecko in 
Firefox, and probably in other browsers, we actually use OCap under the hood 
and have for years. In HTML5, the WindowProxy/Window distinction was finally 
specified, as an ad-hoc instance of OCap membranes.  Any time we deviate from 
OCap, we regret it for both security bug and access-checking overhead reasons.''
\cite{Eich:2013}

We would like a way to declare the authority that an object ought to
possess and then check whether the implementation matches the intent;
that is, we would like a language for declaring security policy.
Drossopoulou and Noble \cite{Drossopoulou:2013:NCP:2489804.2489811}
convincingly argue that none of the current specification methods
adequately capture all of the capability policies required to support
ocap systems. The most intriguing aspect of the capability policies
proposed in \cite{Drossopoulou:2013:NCP:2489804.2489811} is {\em
  deniability}. Deniability differs from usual style of specifications
in the following two aspects: a) it describes policies which are {\em
  open}, i.e. apply to a module as well as all its extensions, and b)
describes {\em necessary} conditions for some effect to take
place. For example, a policy may require that in order for any piece
of code to modify the balance of a bank account, it needs to have
access to that account (access to the account is a necessary
condition), and that this property is satisfied by all extensions of
the account library (open). In contrast, classical Hoare Logic
specifications \cite{Hoare:1969:ABC:363235.363259} describe sufficient
conditions and are closed.  In this paper we concentrate on the use of
the {\rhoc} \cite{DBLP:conf/tgc/MeredithR05} \footnote{the
  \emph{R}eflective \emph{H}igher-\emph{O}rder calculus, not to
  be confused with the $\rho$-calculus
  \cite{CirsteaFaureKirchnerWLRA2004}} for expressing deniability. To
express necessary conditions, we invert the assertion and use
bisimulation (e.g. any piece of code which does not have access to the
account will be bisimilar to code in which the balance of the account
is not modified). To express openness, we adapt the concept of the
adjoint to the separation operator, i.e. any further code P which is
attached to the current code will be bisimilar to P in parallel with
code which does not modify the balance of the account.

In short, we show that the \rhoc\; is an ocap language and sketch a
translation from a subset of JavaScript into the calculus; we also
demonstrate that the corresponding Hennessy-Milner logic suffices to
capture the key notion of deniability.

In the larger scheme, this identifies
\cite{Drossopoulou:2013:NCP:2489804.2489811}'s notion of policy with
the proposition-as-types paradigm, also known as The Curry-Howard isomorphism
\cite{Abramsky:1992:PP:194588.194591}
\cite{Krivine-TheCurryHowardCorre}. The Curry-Howard isomorphism
relates formal logic to type theory; it says that propositions are to
types as proofs are to programs.  Hennessy-Milner logic lets us treat
the type of a concurrent process as an assertion that it belongs to a
set of processes, all of which satisfy some property.  We can treat
these properties as contracts governing the behavior of the process
\cite{Meredith:2003:CT:944217.944236}; in other words, the language of
behavioral types suffices as a security policy language.

\subsubsection{Organization of the rest of the paper}

We introduce the calculus. So equipped we illustrate translating a
simple ocap example into the calculus. Next, we introduce the logic,
and illustrate how we can reason about properties of the translated
program and from there move to interpreting policy, focusing on
deniability. Finally, we conclude with a discussion of some of the
limitations of the approach and how we might approach these in future
work.


\section{The calculus}

Before giving the formal presentation of the calculus and logic where
we interpret ocap programs and policy, respectively, we begin with a
couple of examples that illustrate a design pattern used over and over
in this paper. The first is a mutable single-place cell for storing
and retrieving state.


\begin{lstlisting}[mathescape]
def $Cell( slot, state ) \Rightarrow$ {
  new $( v )$ {
    $v!( state )$
    match {
      $slot ? get( ret ) \Rightarrow$ {
        $v?( s ) \Rightarrow ret!( s )$
        $Cell( slot, s )$
      }
      $slot ? set( s ) \Rightarrow$ { $Cell( slot, s )$ }
    }
  }
}
\end{lstlisting}


We read this as saying that a $Cell$ is parametric in some (initial)
$state$ and a $slot$ for accessing and mutating the cell's state. A
$Cell$ allocates a private channel $v$ where it makes the initially
supplied $state$ available to its internal computations. If on the
channel $slot$ it receives a $get$ message containing a channel $ret$
indicating where to send the state of the cell, it accesses the
private channel $v$ and sends the value it received on to the $ret$
channel; then it resumes behaving as a $Cell$. Alternatively, if it
receives a $set$ message containing some new state $s$, it simply
continues as a $Cell$ instantiated with accessor $slot$ and state $s$.

Despite the fact that this example bears a striking resemblance to code
in any of a number of popular libraries, notably the {\tt AKKA}
library for the {\tt Scala} language, it is only a mildly sugared
form of the direct representation in our calculus. Notice also that
like modern application code, it exhibits encapsulation and separation
of implementation from API: a cell's internal access to
{\tt state} is kept in a private channel {\tt v} while
external access is through {\tt slot}. Another nice thing about
this design pattern is that it scales through composition: when
translating ocap examples expressed in {\tt ECMAScript} to \rhoc\;
we will effectively treat the state of an object as a parallel
composition of cells comprising its state.

The second gadget is an immutable map:

\begin{lstlisting}[mathescape]
def $Map( chan, key_1, state_1, \ldots , key_n, state_n ) \Rightarrow$ {
  new $( v_1, \ldots v_n)$ {
    $v_1!( state_1 )$
    $\ldots$
    $v_n!( state_n )$
    $chan ? get( key_1, ret ) \Rightarrow$ { 
      $v_1?( x ) \Rightarrow ret!( x )$
      $Map( chan, key_1, state_1, \ldots , key_n, state_n )$
    }
    $\ldots$
    $chan ? get( key_n, ret ) \Rightarrow$ { 
      $v_n?( x ) \Rightarrow ret!( x )$
      $Map( chan, key_1, state_1, \ldots , key_n, state_n )$
    }
  }
}
\end{lstlisting}

The design pattern for maps is related to the one for cells, but is
missing the capability to mutate the map. The code does not serve a
request to set any of the values it contains. With these two patterns
in hand it will be much easier to treat the examples. Moreover, as we
will see in the sequel, even with these basic patterns we can already
reason about policy.  The shape of the reasoning about policy for these
examples scales compositionally to reasoning about application-level
policy.

\subsubsection{The \rhoc}

The {\rhoc} \cite{DBLP:journals/entcs/MeredithR05}
is a variant of the asynchronous polyadic {\pic}.  When names are
polarized \cite{conf/fsttcs/Odersky95}, the {\pic} is an ocap
language.  Pierce and Turner \cite{PierceTurner:PictDesign}
defined the programming language Pict as sugar over the polarized {\pic}
together with some supporting libraries; unfortunately, the authority
provided by the libraries violated the principle of least authority.
Ko\u{s}\'{i}k refactored the libraries to create Tamed Pict, recovering
an ocap language, then used it to write a defensively consistent
operating system \cite{kosik:dep}. The {\rhoc} differs from the 
{\pic} in that names are quoted processes
rather than generated by the $\nu$ operator.  Both freshness and 
polarization are provided through the use of namespaces at the type level.

We let ${P, Q, R}$ range over processes and ${x, y, z}$ range over
names, and $\vec{x}$ for sequences of names, $|\vec{x}|$ for the
length of $\vec{x}$, and $\{ \vec{y} / \vec{x} \}$ as partial maps from
names to names that may be uniquely extended to maps from processes to
processes \cite{DBLP:journals/entcs/MeredithR05}.


\begin{grammar}
{M, N} \bc \pzero & \mbox{stopped process} \\
       \;\;\; \bm \; {x}{?}{( y_1, \ldots, y_N )} \Rightarrow {P} & \mbox{input} \\
       \;\;\; \bm \; {x}{!}{( Q_1, \ldots, Q_N )} & \mbox{output} \\
       \;\;\; \bm \; {M}{+}{N} & \mbox{choice} \\
{ P, Q } \bc M & \mbox{include IO processes} \\                                
       \;\;\; \bm \; {P} \juxtap {Q} & \mbox{parallel} \\                                
       \;\;\; \bm \; {*}{x} & \mbox{dereference} \\
{x, y} \bc {@}{P} & \mbox{reference} \\
\end{grammar}


The examples from the previous section use mild (and entirely
standard) syntactic sugar: \texttt{def} making recursive definitions a
little more convenient than their higher-order encodings, \texttt{new}
making fresh channel allocation a little more convenient and
\texttt{match} for purely input-guarded choice. Additionally, the
examples use $\{\}$-enclosed blocks together with line breaks, rather
than $\binpar{}{}$ for parallel composition. Thus the expression $v!(
state )$ is actually a thread running in parallel with the
\texttt{match} expression in the $Cell$ definition. The interested
reader is directed to the appendix for more details.

\subsection{Free and bound names}

The syntax has been chosen so that a binding occurrence of a name is
sandwiched between round braces, ${(} \cdot {)}$. Thus, the
calculation of the free names of a process, $P$, denoted
$\freenames{P}$ is given recursively by

\begin{equation*}
  \begin{aligned}
    & \freenames{\pzero} \defneqls \emptyset \\
    & \freenames{{x}{?}{( y_1, \ldots, y_N )} \Rightarrow {P}} \defneqls \\
    & \;\;\;\;\;\{ x \} \cup (\freenames{P} \setminus \{ y_1, \ldots y_N \}) \\
    & \freenames{{x}{!}{( Q_1, \ldots, Q_N )}} \defneqls \{ x \} \cup \bigcup \freenames{Q_i} \\
    & \freenames{\binpar{P}{Q}} \defneqls \freenames{P} \cup \freenames{Q} \\
    & \freenames{{*}{x}} x\defneqls \{ x \} \\
  \end{aligned}
\end{equation*}

An occurrence of $x$ in a process $P$ is \textit{bound} if it is not
free. The set of names occurring in a process (bound or free) is
denoted by $\names{P}$.

\subsection{Structural congruence}

The {\em structural congruence} of processes, noted $\scong$, is the
least congruence containing $\alpha$-equivalence, $\alphaeq$, that
satisfies the following laws:

\begin{eqnarray*}
	{P} \juxtap \pzero	
		&  \scong \; {P} \; \scong & 
			\pzero \juxtap {P} \\
	{P} \juxtap {Q}	
		& \scong & 
			{Q} \juxtap {P} \\
	({P} \juxtap {Q}) \juxtap {R}
		& \scong & 
			{P} \juxtap ({Q} \juxtap {R}) \\
\end{eqnarray*}

\paragraph{Name equivalence} As discussed in
\cite{DBLP:conf/tgc/MeredithR05} the calculus uses an equality,
$\nameeq$, pronounced name-equivalence, recursively related to
$\alpha$-equivalence and structural equivalence, to judge when two
names are equivalent, when deciding synchronization and substitution.

\subsection{Semantic substitution}

The engine of computation in this calculus is the interaction between
synchronization and a form of substitution, called semantic
substitution. Semantic substitution differs from ordinary syntactic
substitution in its application to a dropped name. For more details
see \cite{DBLP:journals/entcs/MeredithR05}

\begin{eqnarray*}
(\dropn{x})  \psubstp{Q}{P}       
		& \defneqls & 
		\left\{ 
			\begin{array}{ccc} 
				Q & & x \nameeq \quotep{P} \\
                              	\dropn{x} & & otherwise \\
			\end{array}
		\right.
\end{eqnarray*}

Finally equipped with these standard features we can present the
dynamics of the calculus.

\subsection{Operational Semantics}
The reduction rules for {\rhoc}  are


\infrule[Comm]
{ {x}_{0} \nameeq {x}_{1}, |\vec{y}| = |\vec{Q}| }
{{{ x_{0}{?}{(}{\vec{y}}{)} \concat {P}}\juxtap {x_{1}}{!}{(}{\vec{Q}}{)}}
\red {{P}{\{}\quotep{\vec{Q}}{/}{\vec{y}}{\}}}}

In addition, we have the following context rules:

\infrule[Par]{{P} \red {P}'}{{{P} \juxtap {Q}} \red {{P}' \juxtap {Q}}}

\infrule[Equiv]{{{P} \scong {P}'} \andalso {{P}' \red {Q}'} \andalso {{Q}' \scong {Q}}}{{P} \red {Q}}

The context rules are entirely standard and we do not say much about
them here. The communication rule makes it possible for agents to
synchronize at name-equivalent guards and communicate processes
packaged as names. Here is a simple example of the use of
synchronization, reduction and quasiquote: $x{(}{z}{)} \Rightarrow
{w}{!}{(}{y}{!}{(}{*}{z}{)}{)} \juxtap x{!}{(}{P}{)} \red
{w}{!}{(}{y}{!}{(}P{)}{)}$. The input-guarded process, $x{(}{z}{)}
\Rightarrow {w}{!}{(}{y}{!}{(}{*}{z}{)}{)}$, waits to receive the code
of $P$ from the output-guarded data, $x{!}{(}{P}{)}$, and then reduces
to a new process the body of which is rewritten to include $P$,
${w}{!}{(}{y}{!}{(}P{)}{)}$.

\paragraph{Bisimulation}

One of the central features of process calculi, referred to in the
literature as bisimulation, is an effective notion of
substitutability, i.e. when one process can be substituted for
another, and with it a range of powerful proof techniques. For this
paper we focus principally on the use of logic to interpret policy and
so move a discussion of bisimulation to an appendix. The main theorem
of the next section relates the logic to bisimulation. The reader
unfamiliar with this important aspect of programming language
semantics is strongly encouraged to look at
\cite{DBLP:LNCS/Sangiorgi06}.

\section{Interpreting capabilities}

Rather than giving a translation of the whole of {\tt ECMAScript} into
\rhoc, we focus on one example from
\cite{Drossopoulou:2013:NCP:2489804.2489811} designed to illustrate
key aspects of object-capabilities and requirements for a policy
language.

\begin{verbatim}
var makeMint = () => {
  var m = WeakMap();
  var makePurse = () => mint(0);
  var mint = balance => {
    var purse = def({
      getBalance: () => balance,
      makePurse: makePurse,
      deposit:
       (amount, srcP) => Q(srcP).then(src => {
         Nat(balance + amount);
         m.get(src)(Nat(amount));
         balance += amount;
       })
    });
    var decr = amount => {
      balance = Nat(balance - amount);
    };
    m.set(purse, decr);
    return purse;
  };
  return mint;
};
\end{verbatim}

Interpreting this code illustrates how surprisingly intuitive and
natural it is to represent and reason about object-capabilities in
\rhoc. Essentially, a \\ capability---{\em i.e.} access to some behavior, power
or authority---is represented by a channel. When an agent in the \rhoc\;
has access to a channel it has exactly that: access to some behavior,
power or authority.


Armed with $Cell$ and $Map$ from the first section, we can translate
each of the language elements from the bank example into \rhoc.  Note
that the interpretation takes a channel $k$ on which to signal
completion of the statement; synchronous JavaScript necessarily needs
to be transformed into continuation-passing style as part of the
interpretation in {\rhoc}.  The asynchronous {\tt Q} library used in
the definition of the \texttt{deposit} method, on the other hand, is
unneeded because {\rhoc} is inherently asynchronous.
\begin{equation*}
  \begin{aligned}
    & \meaningof{\texttt{var}\; x; P}( x, k ) \defneqls \binpar{Cell( x, \texttt{undefined} )}{\meaningof{P}( x, k )} \\
    & \meaningof{\texttt{var}\; x = v; P}( x, k ) \defneqls \\
    & \;\;\;\;\;\texttt{new}(k_1)\{ \binpar{Cell( x, \meaningof{ v }(x, k_1 ) )}{k_1?() \Rightarrow \meaningof{P}( x, k )}\} \\
    & \meaningof{ P ; Q }( x, k ) \defneqls \texttt{new}(k_1)\{ \binpar{\meaningof{ P }( x, k_1 )}{ k_1?() => \meaningof{ Q }( x, k )} \} \\
    & \meaningof{ \texttt{i} +\!= \texttt{a} ; P }( x, k ) \defneqls \\
    & \;\;\;\;\;\texttt{new}(k_1) \{ \\
    & \;\;\;\;\;\;\;\;\texttt{i} ! get( r ) \\
    & \;\;\;\;\;\;\;\;{r?( v ) \Rightarrow \texttt{\{} \binpar{\texttt{i}! set( v + \meaningof{\texttt{a}}( x, k ) )}{k_1!()} \texttt{\}} } \\
    & \;\;\;\;\;\;\;\;{\meaningof{P}( x, k_1 )} \\
    & \;\;\;\;\; \} \\
    & \meaningof{\texttt{def}(\{key_1: state_1, \ldots, key_n: state_n\})}(x, k) \defneqls \\
    & \;\;\;\;\;\binpar{ Map(x, key_1, state_1, \ldots, key_n, state_n)}{k!()} \\
  \end{aligned}
\end{equation*}

We do not give an explicit translation of {\tt WeakMap} because it is quite
large, but it is essentially the same as $Map$ plus some machinery for performing
updates.



The key point is that the translation is compositional, so we can
build up the translation of the whole from the translation of the
parts. Moreover, we can reason equationally in the translated
code. For example, to translate the code fragment below we calculate
directly


\begin{lstlisting}[mathescape]
$\ldb \mbox{\tt var}\; balance = initAmt;$
$\; \; \mbox{\\} balance\; +\!= \; amount; P \rdb( x, k )$
$=$ (defn)
{ 
  ${Cell( balance, \meaningof{initAmt}( x, k ) )}$
  ${k!()}$
  new $( rslt, k_1 )$ { 
    ${balance ! get( rslt )}$
    $rslt?( b ) \Rightarrow$ {
      $balance ! set( b + \meaningof{ amount }( x, k ) )$
      $k_1!()$
    }
    ${k_1?() \Rightarrow \meaningof{ P }( x, k )}$
  }
}
$=$ (extrusion laws for new)
new $( rslt, k_1 )$ {
  ${Cell( balance, \meaningof{initAmt}( x, k ) )}$
  ${k!()}$
  $balance ! get( rslt )$
  $rslt?( b ) \Rightarrow ${
    $balance ! set( b + \meaningof{amount}( x, k ) )$
    $k_1!() $
  }
  ${k_1?() \Rightarrow \meaningof{ P }( x, k )}$
}
\end{lstlisting}

\section{Logic}
Namespace logic resides in the subfamily of Hennessy-Milner logics
known as spatial logics, discovered by Caires and Cardelli
\cite{DBLP:conf/fossacs/Caires04}. Thus, in addition to the action
modalities, we also find at the logical level formulae for
\emph{separation} corresponding to the structural content of the
parallel operator at the level of the calculus. Likewise, we have
quantification over names. There are important differences between the
logic presented here and Caires' logic. The interested reader is
referred to \cite{DBLP:conf/tgc/MeredithR05}.


\subsection{Examples}

A principal advantage to using logical formulae of this kind is that
they denote classes or programs. Unlike the usual notion of type,
where a type's inhabitants are instances of data structures, in
namespace logic a formula (i.e. a \emph{type}) $\phi$ is inhabited by
programs---namely all the programs that satisfy the
formulae. Namespace logic takes this a step further. Because it is
built on a reflective programming language (the \rhoc) it also has
formulae, $\pquotep{\phi}$, for describing classes of names. Said
another way, the inhabitants of types of the form $\pquotep{\phi}$ are
names and hence these types have the right to be called
namespaces. The next two examples show how useful namespaces are in
the security setting.

\subsubsection{Controlling access to namespaces}
\label{namespace}
Suppose that $\pquotep{\phi}$ describes some namespace, {\em i.e.} some
collection of names. We can insist that a process restrict its next
input to names in that namespace by insisting that it witness the formula

\begin{eqnarray}
  \pprefix{\pquotep{\phi}}{b}{\ptrue} \& \neg \pprefix{\pquotep{\neg \phi}}{b}{\ptrue} \nonumber
\end{eqnarray}

\noindent which simply says the the process is currently able to take input from
a name in the namespace $\pquotep{\phi}$ and is not capable of input on
any name not in that namespace. In a similar manner, we can limit a
server to serving only inputs in $\pquotep{\phi}$ throughout the
lifetime of its behavior \footnote{Of course, this formula also says
the server never goes down, either---or at least is always willing
to take such input.}

\begin{eqnarray}
  \pgfp{X}{\pprefix{\pquotep{\phi}}{b}{X} \& \neg \pprefix{\pquotep{\neg \phi}}{b}{X}} \nonumber
\end{eqnarray} 

This formula is reminiscent of the functionality of a firewall, except
that it is a \emph{static} check. A process witnessing this formula
will behave as though it were behind a firewall admitting only access
to the ports in $\pquotep{\phi}$ without the need for the additional
overhead of the watchdog machinery.

We will use refinements this basic technique of walling off a behavior
behind a namespace when reasoning about the ocap examples and
especially in the context of deniability.

\begin{grammar}
{\phi, \psi} \bc \ptrue & \mbox{verity} \\
\;\;\; \bm \; \pzero & \mbox{nullity} \\
\;\;\; \bm \; \neg \phi & \mbox{negation} \\
\;\;\; \bm \; \phi \& \psi & \mbox{conjunction} \\
\;\;\; \bm \; \phi \juxtap \psi & \mbox{separation} \\
\;\;\; \bm \; \pdropf{b} & \mbox{disclosure} \\
\;\;\; \bm \; \plift{a}{\phi} & \mbox{dissemination} \\
\;\;\; \bm \; \pprefix{a}{b}{\phi} & \mbox{reception} \\
\;\;\; \bm \; \pgfp{X}{\phi} & \mbox{greatest fix point} \\
\;\;\; \bm \; \pquant{n}{\psi}{\phi} & \mbox{quantification} \\
{a} \bc \pquotep{\phi} & \mbox{demarcation} \\
\;\;\; \bm \; b & \mbox{...} \\
{b} \bc \pquotep{P} & \mbox{nomination} \\
\;\;\; \bm \; n & \mbox{...} \\
\end{grammar}

We let $\PFormula$ denote the set of formulae generated by the
$\phi$-production, $\QFormula$ denote the set of formulae generated by
the $a$-production and $\PropVar$ denote the set of propositional
variables used in the $\textsf{rec}$ production. Additionally, we let
$\PFormula_{-Par}$ denote the separation-free fragment of formulae.



Inspired by Caires' presentation of spatial logic
\cite{DBLP:conf/fossacs/Caires04}, we give the semantics in terms of
sets of processes (and names). We need the notion of a valuation $v :
\PropVar \to \wp(\Proc)$, and use the notation $v\substn{\mathcal{S}}{X}$ to mean 

\begin{eqnarray}
  v\substn{\mathcal{S}}{X}(Y) & = &
  \left\{ \begin{array}{ccc}
      S & & Y = X \\
      v(Y) & & otherwise \\
    \end{array}
  \right.\nonumber
\end{eqnarray}

The meaning of formulae is given in terms of two mutually recursive functions,

\begin{eqnarray}
\pmeaningof{ - }( - ) : \PFormula \times [\PropVar \to \wp(\Proc)] \to \wp(\Proc) \nonumber\\
\nmeaningof{ - }( - ) : \QFormula \times [\PropVar \to \wp(\Proc)] \to \wp(\QProc) \nonumber
\end{eqnarray}

\noindent taking a formula of the appropriate type and a valuation, and
returning a set of processes or a set of names, respectively.

\begin{equation*}
  \begin{aligned}
  & \pmeaningof{\ptrue}(v) \defneqls \Proc \\ 
  & \pmeaningof{\pzero}(v) \defneqls \{ P : P \scong \pzero \}  \\ 
  & \pmeaningof{\neg \phi}(v) \defneqls \Proc / \pmeaningof{\phi}(v) \\
  & \pmeaningof{\phi \& \psi}(v) \defneqls \pmeaningof{\phi}(v) \cap \pmeaningof{\psi}(v) \\
  & \pmeaningof{\binpar{\phi}{\psi}}(v) \defneqls \\
  & \;\;\;\;\;\{ P : \exists P_0, P_1.P \scong \binpar{P_0}{P_1}, \; P_0 \in \pmeaningof{\phi}(v), \;  P_1 \in \pmeaningof{\psi}(v) \} \\
  & \pmeaningof{\pdropf{b}}(v) \defneqls \{ P : \exists x.P \scong {\dropn{x}}, \; x \in \nmeaningof{b}(v) \} \\	
  & \pmeaningof{\plift{a}{\phi}}(v) \\
  & \defneqls \{ P : \exists P'.P \scong {\lift{x}{P'}},
                                           \; x \in \nmeaningof{a}(v), 
                                           \; P' \in \pmeaningof{\phi}(v) \} \\
  & \pmeaningof{\pprefix{a}{b}{\phi}}(v) \defneqls \\
  & \;\;\;\;\;\{ P : \exists P'.P \scong {\prefix{x}{y}{P'}}, x \in \nmeaningof{a}(v), \\
  & \;\;\;\;\;\; \; \; \forall c . \exists z . {P'}\substn{z}{y} \in \pmeaningof{{\phi}\substn{c}{b}}(v) \} \\
  & \pmeaningof{\pgfp{X}{\phi}}(v) \defneqls \cup \{ \mathcal{S} \subseteq \Proc : \mathcal{S} \subseteq \pmeaningof{\phi}(v\substn{\mathcal{S}}{X})\} \\
  & \pmeaningof{\pquant{n}{\psi}{\phi}}(v) \defneqls \cap_{x \in \nmeaningof{\quotep{\psi}}(v)} \pmeaningof{{\phi}\substn{x}{n}}(v) \\
  & \nmeaningof{\pquotep{\phi}}(v) \defneqls \{ x : x \nameeq \quotep{P}, P \in \pmeaningof{\phi}(v) \} \\
  & \nmeaningof{\pquotep{P}}(v) \defneqls \{ x : x \nameeq  \quotep{P} \} \\
  \end{aligned}
\end{equation*}

We say $P$ witnesses $\phi$ (resp., $x$ witnesses $\pquotep{\phi}$),
written $P \models \phi$ (resp., $x \models \pquotep{\phi}$) just when
$\forall v . P \in \meaningof{\phi}(v)$ (resp., $\forall v . x \in \meaningof{\pquotep{\phi}}(v)$).

\begin{thm}[Equivalence]\label{sec:equivalence_theorem}
	$P \wbbisim Q \riff \forall \phi \in \PFormula_{-Par} . P \models \phi \riff Q \models \phi .$
\end{thm}

The proof employs an adaptation of the standard strategy. A
consequence of this theorem is that there is no algorithm guaranteeing
that a check for the witness relation will terminate. However there is
a large class of interesting formulae for which the witness check does
terminate, see \cite{DBLP:conf/fossacs/Caires04}. {\em Nota bene}: including
separation makes the logic strictly more observant than bismulation.

\subsubsection{Syntactic sugar }

In the examples below, we freely employ the usual DeMorgan-based
syntactic sugar. For example, $\phi \Rightarrow \psi \defneqls \neg (
\phi \& \neg \psi )$ and $\phi \vee \psi \defneqls \neg ( \neg \phi \&
\neg \psi )$. Also, when quantification ranges over all of $\QProc$,
as in $\pquant{n}{\pquotep{\ptrue}}{\phi}$, we omit the typing for the
quantification variable, writing $\pquantuntyped{n}{\phi}$.

\section{Interpreting policy}

Drossopoulou and Noble \cite{Drossopoulou:2013:NCP:2489804.2489811} argue for
several different types of policy specifications. In this paper we
focus on deniability policies, since the others have ready
translations in this setting. Deniability is the one type of policy
specification where the translation is not at all straightforward. For
the reader familiar with logics of concurrency the core idea is to use
rely-guarantee combined with the adjunct to spatial separation.

A visual inspection of the code for $Cell$ reveals that if a process
$P$'s names are fresh with respect to $slot$, then $P$ cannot directly affect
any $Cell( slot, state )$. Unfortunately, visual inspection is not
reliable in settings only mildly more complex, as in the bank example
above; Yee reports that a 90-man-hour review of around a hundred lines
of heavily commented Python code by security specialists was not
sufficient to discover three inserted bugs \cite[Section
7]{Yee:EECS-2007-136}.  However, we can use our logic to state and
verify the fact that $P$ cannot affect any {\tt Cell}. Adapting the
firewall formula, let $soleAccess( slot ) \defneqls
\pgfp{X}{\pprefix{slot}{b}{X} \& \neg \pprefix{\neg slot}{b}{X}}$, and
$noAccess( slot ) \defneqls \pgfp{X}{\neg \pprefix{slot}{b}{X}}$; then
we have
\[Cell (slot, state) \models soleAccess( slot )\]
which implies
\[{\mbox{\tt new}(slot)\mbox{\tt \{} Cell(slot, state) \mbox{\tt \}}}
\wbbisim 0.\]
If we have that $P \models noAccess( slot ), \;$ then 
\[\forall \; state . \mbox{\tt new}(slot)\mbox{\tt \{}
\binpar{P}{Cell(slot, state)} \mbox{\tt \}}\wbbisim P.\]
For, by hypothesis, $slot$ cannot be free in $P$, thus
\begin{equation*}
  \begin{aligned}
    & \mbox{\tt new}(slot)\mbox{\tt \{} \binpar{P}{Cell(slot, state)} \mbox{\tt \}}\\
    & \scong \\
    & \binpar{P}{\mbox{\tt new}(slot)\mbox{\tt \{} Cell(slot, state) \mbox{\tt \}}} \\
  \end{aligned}
\end{equation*}
\noindent which means
\[\binpar{P}{\mbox{\tt new}(slot)\mbox{\tt \{} Cell(slot, state)
  \mbox{\tt \}}} \wbbisim \binpar{P}{0} \scong P.\]
More generally, note that objects involve not just one cell, but
many. By restricting those cells to slots that live in namespace $Slot
\defneqls \pquotep{\phi}$ for some $\phi$ we can scale our formula to
separate objects from a given environment. 

For now, we refine the basic idea to illustrate how to model
deniability. The shape of the logical statement above is: if $Q
\models \phi$ then when $P \models \psi$ we have that
$\texttt{new}(x_1,\ldots,x_N)\{ \binpar{P}{Q} \} \wbbisim S$ for some
properties $\phi$ and $\psi$ and some behavior, $S$. In light of
theorem \ref{sec:equivalence_theorem} we can, in principle, transform
this into a statement of the form, ``If $Q \models \phi,$ then when $P
\models \psi$ we have that $\texttt{new}(x_1,\ldots,x_N)\binpar{P}{Q}
\models \omega$ for some properties $\phi$, $\psi$, and $\omega$.'' This
is the shape of our encoding of deniability policies. The key insight
here is to mediate between closed-world and open-world forms of
deniability. In this open-world formulation we do not quantify over
all potential contexts and attackers, but work just with those that
can be relied upon to use the API, even in attacks.

The first element of this encoding we wish to draw attention to is
that this provides just the emphasis on necessary rather than
sufficient conditions. Moreover, notice that the form of the statement
remains true to the intent of deniability even if the logic is
\emph{more} discriminating than bisimulation---which is the case if we
include the separation operator (as noted in
\cite{DBLP:conf/fossacs/Caires04}). In the presence of a logic with
separation this shape turns out to be a form of rely-guarantee
identified by Honda in \cite{Honda:2008:UTP:2227536.2227558} and, as
Honda observes, can be internalized in the logic as the adjunct to
separation. Thus, we include a form of rely-guarantee formulae, $\phi
\rhd_{\vec{x}} \psi$, where $\vec{x} = \{ x_1,\ldots,x_N\}$, with
semantics $\pmeaningof{\phi \rhd_{\vec{x}} \psi}( v ) = \{ P : \forall
Q. Q \in \pmeaningof{\phi}( v ) \Rightarrow
\mbox{\texttt{new}}(x_1,\ldots,x_N)\mbox{\{}\binpar{P}{Q}\mbox{\}}\in
\pmeaningof{\psi}(v) \}$, as the logical form that mediates between
closed-world and open-world formulations of deniability.

To apply this to the bank example amounts to recognizing that the
\texttt{purse} is essentially a generalized $Cell$. There are two
means to access the internal state: \texttt{deposit} and
\texttt{decr}. The example, however, enjoys an additional level of
indirection in that in the \texttt{ECMAScript} the corresponding
functions these are scoped just to that purse and affect no
other. Attempting to model this naively will quickly run into
problems, e.g.  as a parallel composition of purses will contend over
serving balance update messages. This is where the higher order nature
of the {\rhoc} makes it more suitable than the {\pic} for modeling
these kinds of programs. In our translation a reference to
\texttt{purse} $\meaningof{\texttt{var}\; purse = \ldots}$ is a $Cell$
that holds the \emph{code} of a process for a (generalized) $Cell$
serving the \texttt{deposit} and \texttt{decr} messages. Updates
through a purse first access and execute the behavior to serve balance
update messages and then put the modified \emph{behavior} (with the
new balance) back in the $Cell$ for the purse. Thus, to express the
property that no one can affect the balance of a purse they don't have
\cite{Drossopoulou:2013:NCP:2489804.2489811} we only have to reason
about access through this one channel, $purse$, providing access to
the outer $Cell$. The higher-order nature of the calculus does the
rest of the state management for us. But, we have already accomplished
that for $Cell$s, generically, in our previous reasoning.

\section{Conclusions and future work}

The bulk of our work in modeling deniability policies as
rely-guarantee formulae in a logic with separation was in setting up
the calculi and logic. Once these are absorbed, the actual model is
short and to the point. We take this as evidence that the approach is
natural and commensurate with the expressiveness demands of
deniability style policies.

In this connection it is important to note that a lot is already known
about the complexity characteristics of spatial-behavioral
logics. Interpreting ocap policy in this setting, therefore, has the
further advantage that we are able to classify policy statements that
are potentially too hard to check automatically, or even check at
all. To wit, \cite{Caires04eliminationof} note that the addition of
the adjunct to separation is not a conservative extension!

In developing this approach we uncovered several situations where
value-dependent policies were of interest or importance. This suggests
that some sort of dependent types may be necessary for addressing more
realistic examples.

Finally, source code for implementations of an interpreter for the
{\rhoc} and a model-checker for namespace logic are available upon
request and a more industrial scale version of the core concepts are
used heavily in the \texttt{SpecialK} library available on
\texttt{github} \cite{SpecialK}. This library is already in use in
industrial projects such as the Protunity Business Matching Network
\cite{Protunity}.

\paragraph{Acknowledgments.}
The authors wish to thank Fred Fisher, Mark Miller and James Noble for thoughtful
and stimulating conversation about policy, object capabilities and
types. The first two authors' work on this topic is generously supported
by Agent Technologies, Inc.

\bibliographystyle{amsplain}
\bibliography{tocappi}


\section{Appendix: $\rho$-calculus sugar}

It is known that replication (and hence recursion) can be implemented
in a higher-order process algebra \cite{SangiorgiWalker}. This
encoding provides a good example of calculation with the {\rhoc}.


\begin{equation*}
  \begin{aligned}
    & D(x) \defneqls \prefix{x}{y}{(\binpar{\outputp{x}{y}}{\dropn{y}})} \\
    & \xbangp{P}{x} \defneqls \binpar{\lift{x}{\binpar{D(x)}{P}}}{D(x)} \\
  \end{aligned}
\end{equation*}

\begin{equation*}
  \begin{aligned}
    & \xbangp{P}{x} \\
    & = \\
    & \lift{x}{(\prefix{x}{y}{(\outputp{x}{y} \juxtap \dropn{y})) \juxtap P}} 
    \juxtap \prefix{x}{y}{(\outputp{x}{y} \juxtap \dropn{y})} \\
    & \red  \\
    & (\outputp{x}{y} \juxtap \dropn{y})\substn{\quotep{(\prefix{x}{y}{(\dropn{y} \juxtap \outputp{x}{y})) \juxtap P}}}{y} \\
    & = \\
    & \outputp{x}{\quotep{(\prefix{x}{y}{(\outputp{x}{y} \juxtap \dropn{y})) \juxtap P}}} \\
    & \;\;\;\;\;\juxtap {(\prefix{x}{y}{(\outputp{x}{y} \juxtap \dropn{y})) \juxtap P}} \\
    & \red \\
    & \ldots \\
    & \red^* \\
    & P \juxtap P \juxtap \ldots & \\
  \end{aligned}
\end{equation*}

Of course, this encoding, as an implementation, runs away, unfolding
$\xbangp{P}{x}$ eagerly. A lazier and more implementable replication
operator, restricted to input-guarded processes, may be obtained as follows.

\begin{equation*}
\bangxp{\prefix{u}{v}{P}}{x}
	\defneqls 
	\binpar{\lift{x}{\prefix{u}{v}{(\binpar{D(x)}{P})}}}{D(x)}
\end{equation*}

It is worth noting that the combination of reference and dereference
operators is essential to get computational completeness. 

Armed with this we can consider a conservative extension of the
calculus presented above.

\begin{grammar}
{ P, Q } \bc \ldots \\
\;\;\; \bm \; (\texttt{def} {X}{( y_1, \ldots, y_N )} \Rightarrow {P}){( Q_1, \ldots, Q_N )} \\ 
\;\;\; \bm \; {X}{( Q_1, \ldots, Q_N )} \\
\;\;\; \bm \; \texttt{new}( x_1, \ldots, x_N )P \\
\end{grammar}

The notation for recursive definitions is slightly quirky in that it
simultaneously defines and invokes the definition to some arguments --
in order to make the recursive form be a process expression, as
opposed to a separate category of declaration. Comparing this with the
examples above we have simply omitted the application of the recursive
definitions to initial arguments in the interest of brevity.

\begin{equation*}
  \begin{aligned}
    & (\texttt{def} {X}{( y_1, \ldots, y_N )} \Rightarrow {P}){( Q_1, \ldots, Q_N )} \defneqls \\
    & \;\;\;\binpar{\bangxp{(n( X, P ) \Rightarrow P')}{n( X, P )}}{n( X, P )!( Q_1, \ldots, Q_N )} \\  
  \end{aligned}
\end{equation*}

where $P' \defneqls P \{ n( X, P )!( Q_1, \ldots, Q_N )/ X( Q_1,
\ldots, Q_N )\}$, i.e., the body of $P$ in which every occurrence of

$X( Q_1, \ldots, Q_N )$

is replaced with $n( X, P )!( Q_1, \ldots, Q_N
)$, and $n( X, P )$ generates a name from $X$ guaranteed fresh in $P$.

Likewise, there are several translation of the \pic\; $\texttt{new}$
operator into the \rhoc. Rather than take up space in this paper we
simply refer the reader to \cite{DBLP:journals/entcs/MeredithR05}, or
better yet, invite them to come up with a definition to test their
understanding of the calculus.

Additionally, we call out input-guarded only summations using the notation

\begin{lstlisting}[mathescape]
  match { 
   $x_1?( y_{1,1}, \ldots, y_{1,N} ) \Rightarrow P_1 $
   $ \ldots $
   $x_M?(y_{M,1}, \ldots, y_{M,N'} ) \Rightarrow P_M $ 
  }
\end{lstlisting}

\noindent instead of 

$x_1?( y_{1,1}, \ldots, y_{1,N} ) \Rightarrow P_1 + \ldots + x_M?( y_{M,1},\ldots, y_{M,N'} ) \Rightarrow P_M$

In a \texttt{match} context we avail ourselves of a pattern-matching
syntax, such as $x ? get( ret ) \Rightarrow ret!( v )$ and $x ! get( k
)$. Again, encodings are quite standard, e.g. see
\cite{Brown05piduce:a}.

Finally, we use $\{ \ldots \}$ for grouping expressions and in more complex examples, omit
$\binpar{}{}$ in favor of line breaks for legibility. Thus, an expression for a classic race-condition $\binpar{ x?( y_1, \ldots, y_N ) \Rightarrow P}{\binpar{x!( Q_1, \ldots, Q_N )}{ x?( z_1, \ldots, z_N ) \Rightarrow R}}$ might be written
\begin{lstlisting}[mathescape]
  {
    $x?( y_1, ... , y_N ) \Rightarrow P$
    $x!( Q_1, ... , Q_N )$
    $x?( z_1, ... , z_N ) \Rightarrow R$
  }
\end{lstlisting}

\section{Appendix: bisimulation}

Having taken the notion of restriction out of the language, we
carefully place it back into the notion of observation, and hence into
the notion of program equality, {\em i.e.} bisimulation. That is, we
parameterize the notion of barbed bisimulation by a set of names over
which we are allowed to set the barbs. The motivation for this choice
is really comparison with other calculi. The set of names of the
{\rhoc} is \textit{global}. It is impossible, in the grammar of
processes, to guard terms from being placed into contexts that can
potentially observe communication, so we provide a place for
reasoning about such limitations on the scope of observation in the
theory of bisimulation.

From an ocap perspective, having the ability to enumerate names is
usually a sign that a language is not even memory-safe, let alone
capability-safe.  However, in Section \ref{namespace} below, we show
that we can use types to create namespaces and then statically prove
which namespaces a process has access to.


\begin{defn}
An \emph{observation relation}, $\downarrow_{\mathcal N}$, over a set
of names, $\mathcal N$, is the smallest relation satisfying the rules
below.

\infrule[Out-barb]{y \in {\mathcal N}, \; x \nameeq y}
		  {\outputp{x}{v} \downarrow_{\mathcal N} x}
\infrule[Par-barb]{\mbox{$P\downarrow_{\mathcal N} x$ or $Q\downarrow_{\mathcal N} x$}}
		  {\binpar{P}{Q} \downarrow_{\mathcal N} x}

We write $P \Downarrow_{\mathcal N} x$ if there is $Q$ such that 
$P \wred Q$ and $Q \downarrow_{\mathcal N} x$.
\end{defn}

Notice that $\prefix{x}{y}{P}$ has no barb.  Indeed, in {\rhoc} as well
as other asynchronous calculi, an observer has no direct means to
detect if a sent message has been received or not.

\begin{defn}
An  ${\mathcal N}$-\emph{barbed bisimulation} over a set of names, ${\mathcal N}$, is a symmetric binary relation 
${\mathcal S}_{\mathcal N}$ between agents such that $P\rel{S}_{\mathcal N}Q$ implies:
\begin{enumerate}
\item If $P \red P'$ then $Q \wred Q'$ and $P'\rel{S}_{\mathcal N} Q'$.
\item If $P\downarrow_{\mathcal N} x$, then $Q\Downarrow_{\mathcal N} x$.
\end{enumerate}
$P$ is ${\mathcal N}$-barbed bisimilar to $Q$, written
$P \wbbisim_{\mathcal N} Q$, if $P \rel{S}_{\mathcal N} Q$ for some ${\mathcal N}$-barbed bisimulation ${\mathcal S}_{\mathcal N}$.
\end{defn}







\end{document}